\documentclass[twocolumn,aps,prl,preprintnumbers]{revtex4}
\usepackage{}
\usepackage{bbm}
\usepackage[latin9]{inputenc}
\usepackage{amsmath}
\usepackage{amssymb}
\usepackage{graphicx}
\usepackage{mathrsfs}
\usepackage{amsfonts}
\usepackage{amsthm}
\usepackage{color}
\usepackage{txfonts}
\usepackage{lipsum}
\usepackage{gensymb}
\usepackage[colorlinks=true,citecolor=blue,linkcolor=blue,urlcolor=blue,anchorcolor=blue]{hyperref}%
\hypersetup{colorlinks=true,citecolor=blue,linkcolor=blue,urlcolor=blue}
\providecommand{\U}[1]{\protect\rule{.1in}{.1in}}
\makeatletter

\newcommand{\Rmnum}[1]{\expandafter\@slowromancap\romannumeral #1@}
\makeatother
\makeatletter
\@ifundefined{textcolor}{}
{
\definecolor{BLACK}{gray}{0}
\definecolor{WHITE}{gray}{1}
\definecolor{RED}{rgb}{1,0,0}
\definecolor{GREEN}{rgb}{0,1,0}
\definecolor{BLUE}{rgb}{0,0,1}
\definecolor{CYAN}{cmyk}{1,0,0,0}
\definecolor{MAGENTA}{cmyk}{0,1,0,0}
\definecolor{YELLOW}{cmyk}{0,0,1,0}
}
\makeatother
\begin{document}
\title{Generation of twisted magnons via spin-to-orbital angular momentum conversion}
\author{Z.-X. Li}
\author{Zhenyu Wang}
\author{Yunshan Cao}
\author{Peng Yan}
\email[Corresponding author: ]{yan@uestc.edu.cn}
\affiliation{School of Electronic Science and Engineering and State Key Laboratory of Electronic Thin Films and Integrated Devices, University of Electronic Science and Technology of China, Chengdu 610054, China}

\begin{abstract}
Twisted magnons (TMs) carrying orbital angular momentum (OAM) have attracted much attention from the magnonic community. The fabrication of such novel magnon state however is still challenging. Here we present a simple method to generate TMs with arbitrary radial and azimuthal quantum numbers through the spin-to-orbital angular momentum conversion. The conversion rate from plane-wave magnons to twisted ones is shown to be insensitive to the quantum index. The spectrum of TMs in thin nanodisks is solved analytically, showing a good agreement with micromagnetic simulations. Moreover, we numerically study the propagation of TMs in magnetic nanodisk arrays and obtain the quantitative dependence of the decay length on quantum indexes. Our results are helpful for realizing TMs with large OAMs that are indispensable for future high-capacity magnonic communications and computings.
\end{abstract}
\maketitle
Quantized orbital angular momentum (OAM) states of (quasi-)particles have attracted growing interest due to the peculiar twisted phase structure in a broad field of photonics \cite{AllenPRA1992,MairN2001,MolinaNP2007,PadgettOE2017,JiS2020}, electronics \cite{UchidaN2010,VerbeeckN2010,McmorranS2011,SilenkoPRL2017,LloydRMP2017}, acoustics \cite{DashtiPRL2006,AnhauserPRL2017,BareschPRL2018,MarzoPRL2018,BliokhPRB2019}, and neutron sciences \cite{ClarkN2015,CappellettiPRL2018,LarocqueNP2018}. Very recently, the concept of OAM state has been extended to magnons (spin waves)---the elementary excitations in ordered magnets, synthesizing twisted magnons (TMs) \cite{JiangPRL2020,JiaNC2019}. With the explosive growth of the information, how to improve the bit rate of information has become the primary consideration for engineers and scientists. A promising solution is to use multiple channels to deal with such huge amount of information. For example, the distinct spatial profiles of electromagnetic wave (twisted photon) can act as individual information channels, through which one can realize OAM-based spatial-division multiplexing (SDM) \cite{GibsonOE2005,TamburiniNJP2012,WangNP2012,BozinovicS2013,YuanPRA2021}. Magnons can also be used to carry, transmit, and process information \cite{KruglyakJPD2010,SergaJPD2010,LenkPR2011,ChumakNP2015,LiPR2021}. The high-capacity communication has become a key demand for the development of the emerging field of magnonics. Recently, the frequency-division multiplexers/dumultiplexers have been designed to increase the communication ability of conventional magnons \cite{VogtNC2014,SadovnikovAPL2016,HeussnerPSS2018,HeussnerPSS2020,WangNC2021}. However, to the best of our knowledge, the multiplexing of TMs has never been reported. By utilizing the twisted phase structure of TMs with different frequencies, it is possible to realize frequency-division multiplexing (FDM) which can enhance the capacity of communication based on magnons. Although the structures and properties of TMs have received intensive theoretical investigations \cite{ChenAPL2020,JiaJO2019}, the experimental realization of TMs still represents a challenge so far. The main reason is that the excitation of TMs in magnetic nanodisks or nanocylinders often requires a complex profile of the driving fields. It is thus of vital importance to find a simple and effective method to generate TMs.

In this work, we study the generation and propagation of TMs in magnetic nanodisk arrays. By introducing a spin-to-orbit conversion scheme, we propose a simple method to generate arbitrary TMs. We calculate the conversion rate from planar magnons to TMs and show that it is insensitive to the OAM and radial indexes of TMs and reaches a universal value. The propagation length of TMs in nanodisk arrays, a key parameter for FDM, is also analyzed. Our results may represent an essential step for realizing the TM-based high-capacity communication for magnonic spintronics. 

We first consider a single thin magnetic nanodisk with radius $r$ and thickness $d$. The magnetic moments are perpendicularly magnetized by an out-of-plane static magnetic field $H_{0}$. The magnetization dynamics of the nanodisk is governed by the Landau-Lifshitz-Gilbert (LLG) equation \cite{GilbertPR1955}: 
\begin{equation}\label{Eq1}
\frac{\partial \mathbf{m}}{\partial t}=-\gamma\mathbf{m}\times\mathbf{H}_{\text{eff}}+\alpha\mathbf{m}\times\frac{\partial \mathbf{m}}{\partial t},
\end{equation}
where $\gamma$ is the gyromagnetic ratio, $\alpha$ is the Gilbert damping, $\mathbf{m}$ is the unit vector along the local magnetic moment, and
\begin{equation}\label{Eq2}
\mathbf{H}_{\text{eff}}=H_{0}\hat{z}+\mathbf{h}(\mathbf{r},t)+\frac{2A}{M_{s}}\nabla^{2}\mathbf{m}(\mathbf{r},t)-N_{z}\mu_{0}M_{z}\hat{z},
\end{equation}
is the effective field. In the right-hand side of  \eqref{Eq2}, the first term is the external magnetic field, the second term is the dynamic dipolar field, the third term is the exchange field, and the last term is the static demagnetization field, with $A$ the exchange stiffness, $M_{s}$ the saturation magnetization, $N_{z}$ the demagnetizing factor along $z$ axis, and $\mu_{0}$ being the vacuum permeability. We are interested in the thin-film limit, i.e., $d\ll r$, so we have $N_{z}=1$. The dipolar field $\mathbf{h}(\mathbf{r},t)$ must satisfy the Maxwell's equations: $\nabla\times\mathbf{h}(\mathbf{r},t)=0$ and $\nabla\cdot[\mathbf{h}(\mathbf{r},t)+M_{s}\mathbf{m}(\mathbf{r},t)]=0$. We thus have $\mathbf{h}(\mathbf{r},t)=-\nabla\Phi(\mathbf{r},t)$ with $\Phi$ being the magnetostatic potential. 
\begin{figure}[!htbp]
\begin{centering}
\includegraphics[width=0.48\textwidth]{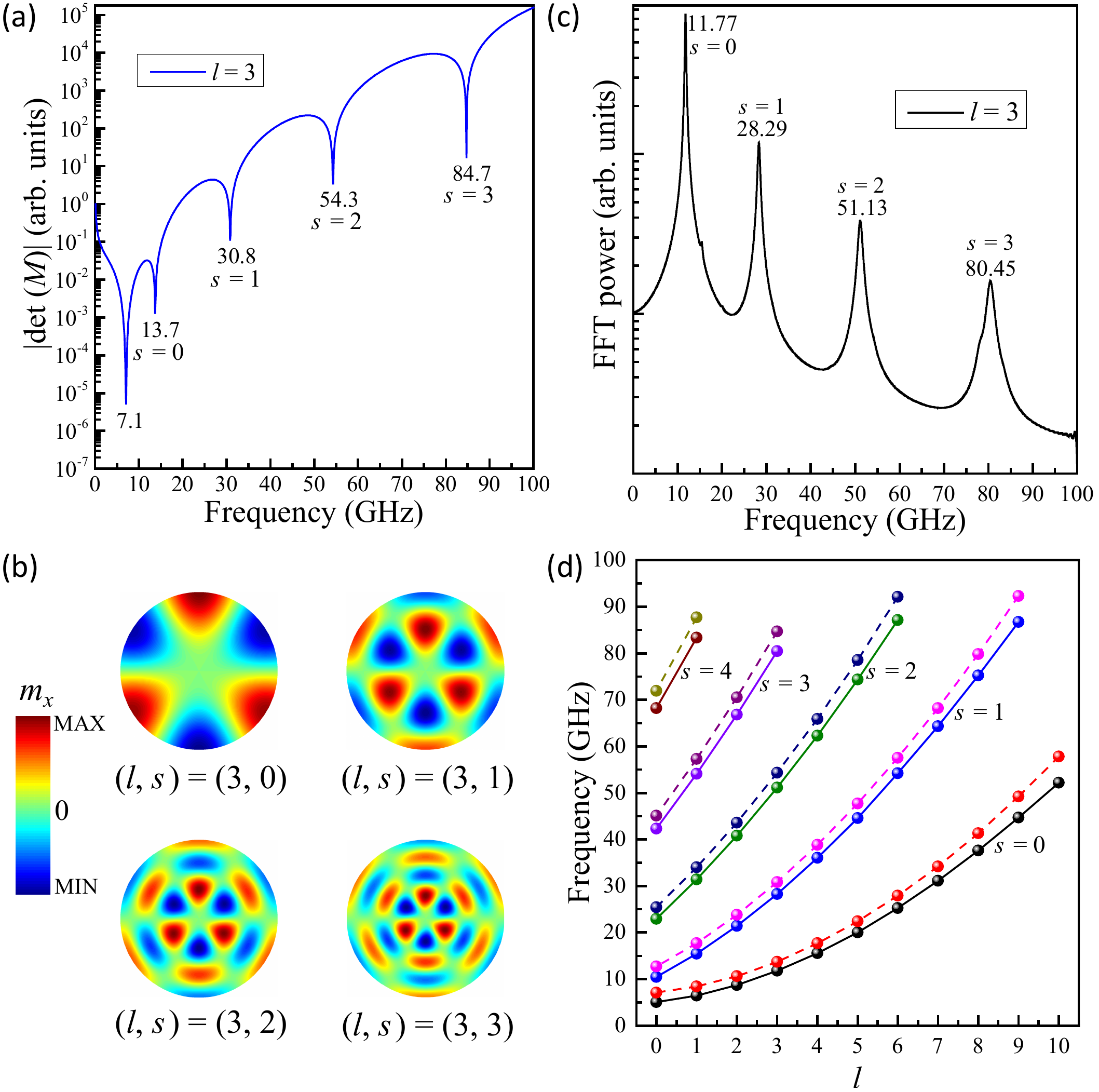}
\par\end{centering}
\caption{(a) The magnitude of the coefficient determinant versus trial frequencies for $l=3$. (b) The spatial distribution of TM modes with different radial quantum number $s=0$, 1, 2, 3 for a fixed OAM quantum number ($l=3$) in the nanodisk. (c) The temporal Fourier spectrum for magnetization component $m_{x}$ of the nanodisk. (d) Dependence of the intrinsic frequencies of TMs on $l$ for different $s$, the dashed and solid lines denote the results from theoretical calculations and micromagnetic simulations, respectively.}
\label{Figure1}
\end{figure}

We then consider the spin wave excitation $\mathbf{m}=(m_{x},m_{y},1)$. By assuming $m_{x(y)}(\mathbf{r},t)=m_{x(y)}(\mathbf{r})e^{-i\omega t}$ and $\Phi(\mathbf{r},t)=\Phi(\mathbf{r})e^{-i\omega t}$, the LLG equation can be simplified as (the damping term and high-order quantities are dropped):
\begin{equation}\label{Eq3}
\begin{aligned}
i\bar{\omega}m_{x}&=(H_{0}-\mu_{0}M_{s}-D\nabla^{2})m_{y}+\frac{\partial\Phi}{\partial y},\\
-i\bar{\omega}m_{y}&=(H_{0}-\mu_{0}M_{s}-D\nabla^{2})m_{x}+\frac{\partial\Phi}{\partial x},\\
\nabla^{2}\Phi&=M_{s}(\frac{\partial m_{x}}{\partial x}+\frac{\partial m_{y}}{\partial y}),
\end{aligned}
\end{equation}
with $\bar{\omega}=\omega/\gamma$ and $D=2A/M_{s}$. Inside the thin nanodisk, the magnetic potential takes the following form:
\begin{equation}\label{Eq4}
\Phi(\rho,\phi)=J_{n}(\kappa\rho)\text{exp}(in\phi),
\end{equation}  
where $J_{n}(\kappa\rho)$ is the Bessel function of the first-kind with order $n$, $\rho=\sqrt{x^2+y^2}$, and $\kappa$ is the transverse wave number. Substituting \eqref{Eq4} into \eqref{Eq3}, we obtain the dispersion relation of the TMs:
\begin{equation}\label{Eq5}
D^{2}\kappa^{4}+(2H_{0}-\mu_{0}M_{s})D\kappa^{2}+[H_{0}(H_{0}-\mu_{0}M_{s})-\bar{\omega}^{2}]=0.
\end{equation}  
This equation is quadratic in $\kappa^{2}$, so that for each $n$, we have two linearly independent solutions. Thus, the magnetic potential within the nanodisk can be expressed as \cite{Limbook2011,AriasPRB2001}:
\begin{equation}\label{Eq6}
\Phi^{\text{disk}}(\rho,\phi)=\sum_{j=1}^{2}C_{j}J_{n}(\kappa_{j}\rho)\text{exp}(in\phi).
\end{equation}  

To obtain the eigenfrequencies of TMs in Eq. \eqref{Eq5}, one needs to determine parameters $\kappa_{j}$ and coefficients $C_{j}$ with $j=1,2$, which can be solved by using the free boundary conditions:
\begin{equation}\label{Eq9}
\Big(\frac{\partial m_{\rho}}{\partial \rho}\Big)\Big|_{\rho=r}=0, \,\,\,\,\,\,\Big(\frac{\partial m_{\phi}}{\partial \rho}\Big)\Big|_{\rho=r}=0.
\end{equation}
Here $m_{\rho}$ and $m_{\phi}$ are the radial and azimuthal components of the dynamical magnetization, respectively. From Eq. \eqref{Eq3}, after some algebra, we obtain the explicit expressions of the magnetization components:
\begin{equation}\label{Eq7}
\begin{aligned}
m_{\rho}=\frac{1}{2}\sum_{j=1}^{2}C_{j}\kappa_{j}\Big[\frac{J_{n+1}(\kappa_{j}\rho)}{\Omega+\bar\omega}-\frac{J_{n-1}(\kappa_{j}\rho)}{\Omega-\bar\omega}\Big]\text{exp}(in\phi),
\end{aligned}
\end{equation}
and 
\begin{equation}\label{Eq8}
\begin{aligned}
m_{\phi}=-\frac{i}{2}\sum_{j=1}^{2}C_{j}\kappa_{j}\Big[\frac{J_{n+1}(\kappa_{j}\rho)}{\Omega+\bar\omega}+\frac{J_{n-1}(\kappa_{j}\rho)}{\Omega-\bar\omega}\Big]\text{exp}(in\phi),
\end{aligned}
\end{equation}
with $\Omega=H_{0}-\mu_{0}M_{s}+D\kappa_{j}^{2}$. Substituting Eqs. \eqref{Eq7} and \eqref{Eq8} into Eq. \eqref{Eq9}, we have:
\begin{equation}\label{Eq10}
\begin{aligned}
P_{n,-}(\kappa_{1}r)C_{1}+P_{n,-}(\kappa_{2}r)C_{2}=0,\\
-iP_{n,+}(\kappa_{1}r)C_{1}-iP_{n,+}(\kappa_{2}r)C_{2}=0,
\end{aligned}
\end{equation}
where
\begin{equation}\label{Eq11}
\begin{aligned}
P_{n,\pm}(\kappa_{j}r)&=\frac{\kappa_{j}^{2}}{4}\Big[\frac{J_{n}(\kappa_{j}r)-J_{n+2}(\kappa_{j}r)}{\Omega+\bar\omega}\pm\frac{J_{n-2}(\kappa_{j}r)-J_{n}(\kappa_{j}r)}{\Omega-\bar\omega}\Big].
\end{aligned}
\end{equation}
The condition for the existence of nontrivial solutions of Eq. \eqref{Eq10} is given by $\text{det}[M(\omega,\kappa_{i})]=0$, where $M(\omega,\kappa_{i})$ is the $2\times2$ coefficient matrix. It is worth noting that the OAM quantum number $l$ and the order of Bessel function $n$ satisfy the relation $l=n-1$ \cite{JiangPRL2020}. 

The eigenfrequencies of TMs in nanodisk can be determined as follow: At first, we fix the parameter $l$ and choose a frequency range (0 to 100 GHz for instance). For every trial frequency, we can calculate the transverse wave number $\kappa$ from Eq. \eqref{Eq5}. After that, we judge whether the determinant $\text{det}[M(\omega,\kappa_{i})]$ is equal to 0 \cite{Footnote1}. If so, this trial frequency is accepted as the eigenfrequency of TMs. 

To justify our theoretical analysis, we consider a yttrium iron garnet (YIG) nanodisk with thickness $d=2$ nm and radius $r=50$ nm. The following material parameters are used: the saturation magnetization $M_{s}=1.92\times10^{5}$ A$\,$m$^{-1}$, the exchange stiffness $A=3.1\times10^{-12}$ J$\,$m$^{-1}$. External static magnetic field $H_{0}=400$ mT is applied along $z$ axis (perpendicular to the nanodisk plane). Figure \ref{Figure1}(a) plots the magnitude of the determinant $\text{det}(M)$ for different trial frequencies with $l=3$, one can clearly see that the curve shows five minimum points. Further, by analyzing the spatial distribution of these modes, we identify four eigenmodes of TMs (the minimum point with frequency $f=7.1$ GHz corresponds to a localized mode due to the boundary effect of nanodisk), as shown in Fig. \ref{Figure1}(b). We adopt $s$ to denote the radial quantum number of twisted magnons. 

\begin{figure}[!htbp]
\begin{centering}
\includegraphics[width=0.48\textwidth]{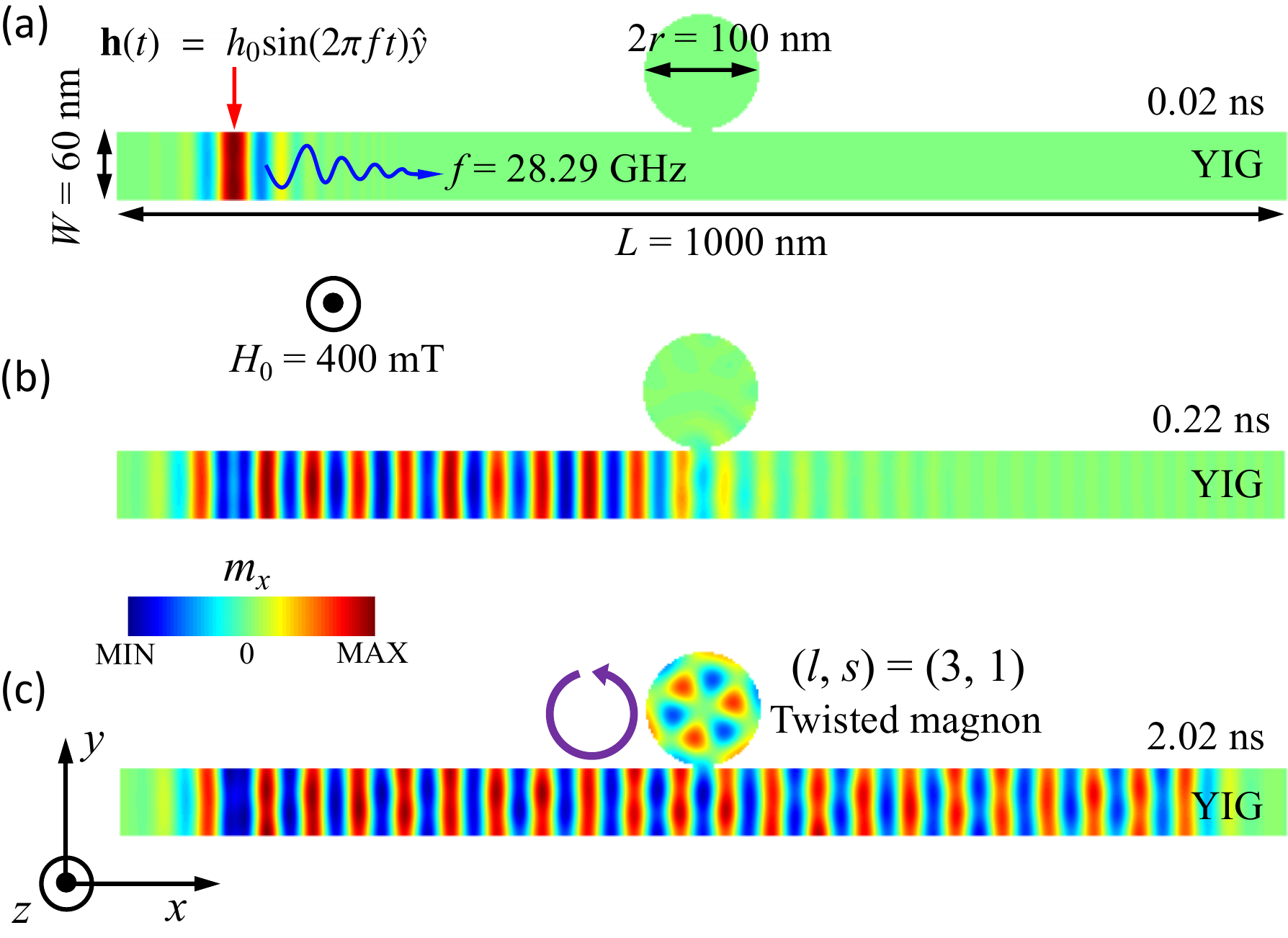}
\par\end{centering}
\caption{Illustration of the generation of TM via the spin-to-orbital angular momentum conversion at $t=0.02$ ns (a), 0.22 ns (b), and 2.02 ns (c). A uniform static magnetic field is applied along the $z$-aixs direction to perpendicularly magnetize the YIG thin film. The red arrow denotes the position of the sinusoidal driving field with frequency $f=28.29$ GHz, and the purple circular arrow represents the rotation direction of TM in the nanodisk.}
\label{Figure2}
\end{figure}     

Micromagnetic simulations are performed to compare with theoretical calculations. The micromagnetic package MUMAX3 \cite{VansteenkisteAA2014} is used to simulate the magnetization dynamics. To generate the TMs, a sinc-function magnetic field
\begin{equation}\label{Eq12}
\mathbf{H}(t)=H_{1}\frac{\text{sin}[2\pi f_0(t-t_{0})]}{2\pi f_0(t-t_{0})}[\text{cos}(l\phi),\text{sin}(l\phi),0]
\end{equation}
is applied to the nanodisk for $100$ ns. Here $H_{1}=10$ mT, $f_0=100$ GHz, $t_{0}=1$ ns, and $l=3$. The temporal fast Fourier transform (FFT) spectrum of the magnetization component $m_{x}$ is plotted in Fig. \ref{Figure1}(c), from which one can clearly see four eigenmodes of TMs. Here we set the Gilbert damping with a large value ($\alpha=0.01$) to speed up the simulation. Similarly, by changing the OAM quantum number $l$, we obtain all eigenfrequencies of TMs within 100 GHz, with the results being shown in Fig. \ref{Figure1}(d). It is noted that the theoretical value (dashed lines) is slightly greater than the micromagnetic result (solid lines), although their comparison is reasonably acceptable. This difference may come from the following reasons: On the one hand, in the theoretical model, we impose the demagnetization factor $N_{z}=1$, while $N_{x}=N_{y}=0$, which is strictly valid only when $d\ll r$. In our calculations, we set $d/r=0.04$ ($d=2$ nm and $r=50$ nm), so there is still room to improve the justification for the approximation. On the other hand, the software MUMAX3 is based on the finite difference method, which may cause errors when handling the curved surfaces. The discrepancy can be reduced by considering nanodisk with a large radius and adopting a smaller mesh size in the simulations.    

\begin{figure}[!htbp]
\begin{centering}
\includegraphics[width=0.48\textwidth]{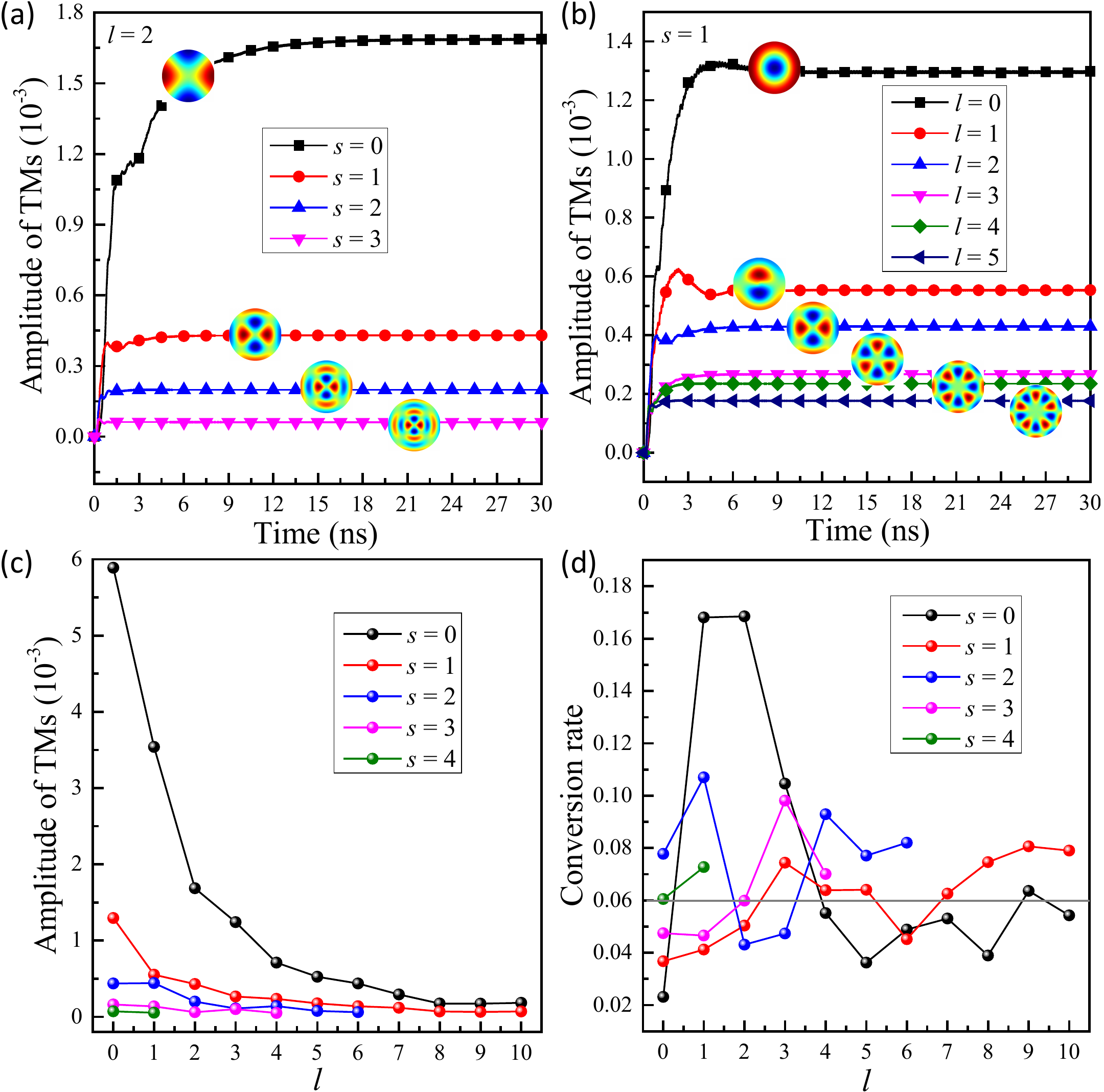}
\par\end{centering}
\caption{Dependence of the TM amplitude on time with different $s$ for a fixed $l$ ($l=2$) (a) and different $l$ for a fixed $s$ ($s=1$) (b). Insets plot the spatial distribution of different TM modes. (c) TM amplitudes under different $l$ and $s$. (d) The conversion rate from planar magnons to TMs varying with $l$ for different $s$. The gray line represent the universal number 6\%.}
\label{Figure3}
\end{figure} 

In the above simulations, we designed exciting fields with very complicated cross sectional structure to generate TMs. The complexity may hinder the practical application of TMs. Here we propose a simple method to generate TMs with any $l$ and $s$. The scheme is plotted in Fig. \ref{Figure2}(a), where the nanodisk contacts a nanostrip of width $W=60$ nm and length $L=1000$ nm. To generate the TMs in nanodisk, we apply a sinusoidal magnetic field $\mathbf{h}(t)=h_{0}\text{sin}(2\pi ft)\hat{y}$ with a specific frequency at one end of the nanostrip (denoted by red arrow) to excite planar magnons, which can be easily realized experimentally within current microstrip antenna technique. As an example, we choose $f=28.29$ GHz, which corresponds to the TM with $l=3$ and $s=1$ according to the dispersion relation [see Fig. \ref{Figure1}(d)]. Besides, we set $h_{0}=1$ mT and damping $\alpha=0.005$ (here we choose a smaller damping to make magnons propagate longer). The absorbing boundaries are adopted at both ends of the nanostrip to prevent the reflection of magnons. When the propagating magnons carrying spin angular momentum pass through the disk-strip touching point, the eigenmode of TM with $(l,s)=(3,1)$ in nanodisk is resonantly excited. Then the planar magnons without any twisting in nanostrip are converted to TMs carrying a finite OAM in nanodisk. We call it a spin-to-orbital angular momentum conversion. In addition, we find that the generated TMs with counterclockwise rotation can stably exist in the nanodisk. The whole process is shown in Fig. \ref{Figure2}. Interestingly, if we place the exciting field to the other end of the nanostrip, the generated TMs will have an opposite rotation direction (clockwise). By setting different frequencies of the local exciting fields, one can generate TMs with any $l$ and $s$.
 
 \begin{figure}[!htbp]
\begin{centering}
\includegraphics[width=0.48\textwidth]{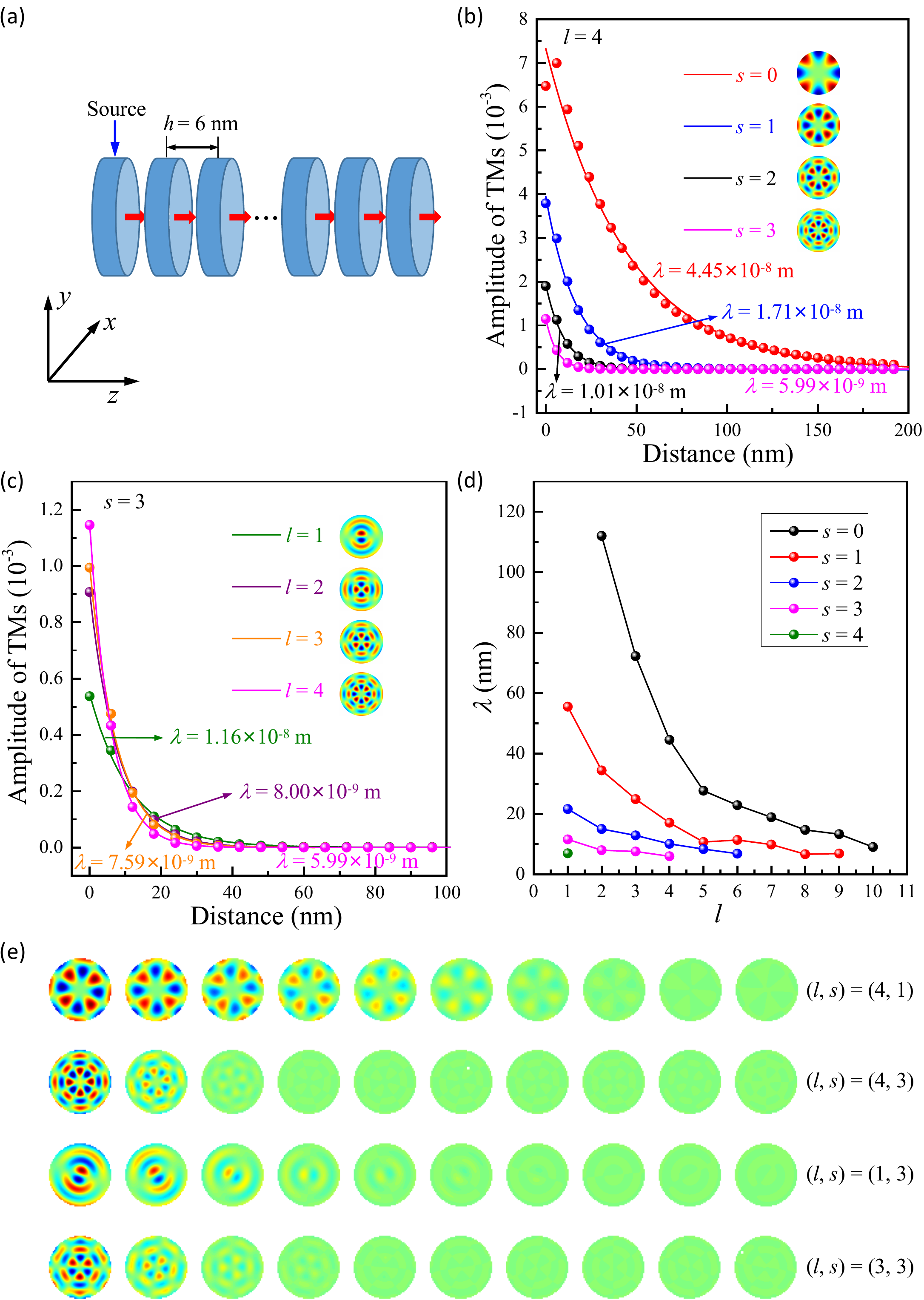}
\par\end{centering}
\caption{(a) The sketch map of a one-dimensional nanodisk array. The blue arrow indicates the position of the applied excitation. Dependence of the TM amplitudes on the propagation distance with different $s$ for a fixed $l$ ($l=4$) (b) and different $l$ for a fixed $s$ ($s=3$) (c). Small balls denote simulation results and solid lines represent the analytical fittings. (d) Dependence of the attenuation length $\lambda$ on $l$ for different $s$. (e) The spatial distribution of TM intensity in nanodisk array for four representative TM modes.}
\label{Figure4}
\end{figure} 
 
To investigate the conversion rate from conventional magnons to TMs in our proposal, we use the quantity $\sqrt{m_{x}^{2}+m_{y}^{2}}$ \cite{WangPRB2012} to quantify the magnon amplitude. Figure \ref{Figure3}(a) plots the amplitude of TMs (averaged over the whole disk) varying with time for different $s$ (here we fixed $l$ to 2). We observe that the intensity of the TMs reaches a stable value very quickly after the exciting field is applied. The larger $s$ we set, the faster the saturation happens. In addition, with the increase of $s$, the intensity of TMs decreases gradually. Similarly, the intensity of TMs decreases with the increase of $l$ with a fixed $s$ (here we set $s=1$ as an example), as shown in Fig. \ref{Figure3}(b). To have a comprehensive picture about the variation trend of the TM intensity, we plot the stable intensity of TMs (after the exciting field is applied for 50 ns) for all different combinations of $s$ and $l$ in Fig. \ref{Figure3}(c), which indeed confirms that the TMs intensity decreases with the increase of $s$ or $l$. The computed conversion rate from conventional magnons to TMs \cite{Footnote2} is plotted in Fig. \ref{Figure3}(d). One can see that the conversion rate reaches a universal number 6\%, although fluctuations exist when quantum indexes $l$ and $s$ vary [see Fig. \ref{Figure2}(c)].

Because of the OAM degree of freedom, TMs have shown outstanding application potentials in future high-capacity communications compared to conventional plane-wave magnons. Here we introduce the concept of magnonic OAM-based FDM in a one-dimensional nanodisk array, with the model shown in Fig. \ref{Figure4}(a). The distance between nearest-neighbor nanodisks is $h$. In the calculations, we set $h=6$ nm if not stated otherwise. When we apply the exciting field of a specific mode frequency in the leftmost nanodisk by using the spin-to-orbit conversion method above, the locally generated TM can propagate due to the magnetostatic interaction between nanodisks. In the linear region, the TMs do not interact with or convert to other TM modes carrying different $l$ or $s$. So we can simultaneously stimulate various TMs carrying different $l$ or $s$, which spread in the waveguide independently, realizing the FDM of TMs. One critical question naturally arises: What's the propagation length of a TM in such one-dimensional nanodisk array?

To address this issue, we study the propagation of TMs with different $s$. As an example, we fix $l=4$. Figure \ref{Figure4}(b) plots the amplitude of TMs varying with propagation distance for different $s$. The small balls are the results of micromagnetic simulations, while the solid lines represent the analytical formula $I=a+be^{-z/\lambda}$. Here $a$ and $b$ are two fitting parameters, $\lambda$ is the propagation length, and $z$ is the distance. We can see that the attenuation of TMs follows the exponential function very well. With the increase of $s$, the intensity of TMs decay even more rapidly. Furthermore, we investigate the propagation of TMs with different $l$, by fixing $s=3$. Simulation results are shown in Fig. \ref{Figure4}(c). It shows that the TMs decay faster with the increase of $l$. The attenuation length $\lambda$ varying with $l$ and $s$ is summarized in Fig. \ref{Figure4}(d), which clearly shows that $\lambda$ decreases with the increase of $l$ or $s$. To illustrate the propagation details of the TMs, we plot the spatial distribution of TM intensity in nanodisk array in Fig. \ref{Figure4}(e), by choosing four representative TM modes. For the sake of observation, we have rotated the angle by 90 degrees. It shows that the OAM carried by TMs is very robust against the dissipation, although the wave amplitude suffers from an obvious decay during the propagation. 

In the present calculations, we have set the radius of nanodisk $r=50$ nm and find almost 40 eigenmodes of TMs below 100 GHz. By choosing a larger radius, we expect to observe more TM modes. In this work, we focused on the propagation property of TMs in nanodisk arrays. The topological property of TMs is also an appealing topic for future study. For example, by constructing the Su-Schrieffer-Heeger \cite{SuPRL1979} and Haldane \cite{HaldanePRL1988} models in disk arrays, one may realize the topological phases of TMs supporting multichannel chiral edge states. Recently, the emergence of higher-order topological phases in ``breathing lattice" have attracted a lot of attention \cite{BenalcazarS2017,HassanNP2018,XueNM2019,ImhofNP2018,Linpj2019}. By constructing a ``breathing" nanodisk array, one may find the so-called higher-order topological TMs. Weyl or Dirac semimetal states of TMs are interesting directions as well. 

To conclude, we have studied the generation and propagation of TMs in magnetic nanodisk arrays. By introducing the concept of spin-to-orbital angular momentum conversion, we propose a simple and effective method to generate TMs with any quantum index. The conversion rate from conventionally planar magnons to TMs is analyzed. We showed that it is insensitive to either the OAM quantum number or the radial index of TMs, but reaches a universal value. We numerically studied the propagation of TMs in magnetic nanodisk arrays and obtained the quantitative dependence of the decay length on quantum indexes, which should be useful for future TM FDM.
\begin{acknowledgments}
We thank Z. Zhang, L. Song, and H. Yang for helpful discussions. This work was supported by the National Natural Science Foundation of China (NSFC) (Grants No. 12074057, No. 11604041, and No. 11704060). Z.-X.L. acknowledges financial support from the NSFC (Grant No. 11904048) and the China Postdoctoral Science Foundation (Grant No. 2019M663461). Z.W. was supported by the China Postdoctoral Science Foundation under Grant No. 2019M653063. 
\end{acknowledgments}


\begin{thebibliography}{99}
\bibitem {Footnote1}Because of the accuracy of numerical calculations, the determinant cannot be strictly equal to $0$. One can identify the zeros by looking for the determinant minimum.
\bibitem {Footnote2}The conversion rate can be calculated as follow: At first, we select two small regions in the nanostrip, which locate on the left and right sides of the disk-strip touching point. Then we calculate the intensity of magnons in this two regions (we can use $P_{1}$ and $P_{2}$ to represent the magnon intensity in the left and right regions, respectively), and $1-P_{2}/P_{1}$ is used to describe the conversion rate.
\end{thebibliography}
\end{document}